\documentclass[lettersize,journal]{IEEEtran}
\usepackage{amsmath,amsfonts}
\usepackage{amssymb}
\usepackage{algorithmic}
\usepackage{array}
\usepackage{balance}
\usepackage[caption=false,font=normalsize,labelfont=sf,textfont=sf]{subfig}
\usepackage{textcomp}
\usepackage{stfloats}
\usepackage{url}
\usepackage{verbatim}
\usepackage{enumitem}
\usepackage{graphicx}
\usepackage{xcolor}
\def\BibTeX{{\rm B\kern-.05em{\sc i\kern-.025em b}\kern-.08em
    T\kern-.1667em\lower.7ex\hbox{E}\kern-.125emX}}
\usepackage{balance}
\begin{document}
\title{A Realistic Coaxial Feed for Cascaded Cylindrical Metasurfaces}

\author{Chun-Wen Lin, \IEEEmembership{Graduate Student Member, IEEE}, and Anthony Grbic, \IEEEmembership{Fellow, IEEE}

\thanks{This work has been submitted to the IEEE for possible publication. Copyright may be transferred without notice, after which this version may no longer be accessible.}
\thanks{This work was supported by the UM-KACST Joint Center for Microwave Sensor Technology. \textit{(Corresponding Author: Anthony Grbic.)}}

\thanks{The authors are with the Radiation Laboratory, Department of Electrical Engineering and Computer Science, University of Michigan, Ann Arbor, MI 48109-2122 USA (email: chunwen@umich.edu; agrbic@umich.edu)}
}

\maketitle

\begin{abstract}
In this letter, a realistic coaxial feed is integrated into the design of cascaded cylindrical metasurfaces. This is in contrast to the fictitious current source that is often reported in literature.
The $S$-matrix of the coaxial feed is obtained by way of the mode-matching technique, which is subsequently combined with the $S$-matrix of the cascaded cylindrical metasurfaces to account for the interaction between the feed and metasurfaces. The integration of a realistic feed into the design process enables practical cylindrical-metasurface-based devices.  
\end{abstract}


\section{Introduction}
\IEEEPARstart{C}{ascaded} cylindrical metasurfaces provide complete control of cylindrical waves \cite{Xu_2020}-\cite{myTAP_2023}. The development of these conformal metasurfaces has enabled angular momentum generation \cite{myTAP_2023}-\cite{Li_2019}, cloaking or illusion \cite{Xu_2020}-\cite{myTAP_2023}, \cite{Kwon_2020}-\cite{Safari_2019}, and high gain antenna design \cite{Xu_2020}-\cite{myTAP_2023}, \cite{Rick_2018}-\cite{Rick_2022}.
Although a few previous works have studied cylindrical metasurfaces fed by antennas \cite{Vellucci_2022}-\cite{Liu_2021}, most literature considered only idealized line current sources as feeds. 
In these scenarios, the interaction between the feed structure and cylindrical metasurfaces has been neglected, which is not realistic.
Hence, it is necessary to introduce a realistic feed and characterize its scattering properties in practical cylindrical metasurface designs.

In \cite{myAPS_2022} we proposed the structure illustrated in Fig. 1. 
A realistic coaxial feed is connected to the center of a radial waveguide, exciting the concentrically cascaded cylindrical metasurfaces. 
In order to characterize the junction between the coaxial feed and radial waveguide, researchers have resorted to analytical derivations involving the Equivalence Principle and image theory \cite{Otto_1967}, or numerical solutions via mode-matching technique \cite{Shen_1999}. 
However, these methods \cite{Otto_1967}-\cite{Shen_1999} have only addressed the azimuthally symmetric case. 
These results are insufficient for the design of azimuthally-varying cylindrical metasurfaces which scatter higher-order azimuthal modes that interact with the coaxial feed.

By applying the mode-matching technique \cite{Shen_1999}-\cite{Eleftheriades_1994} over higher-order azimuthal modes, the scattering properties of the junction for all azimuthal modes of interest have been characterized and summarized into an $S$-matrix \cite{myAPS_2022}. 
In this letter, the mathematical background behind \cite{myAPS_2022} is discussed in detail. 
The coaxial feed is accounted for in the cylindrical metasurface design by integrating the $S$-matrix of the junction into the multimodal wave matrix theory \cite{myTAP_2023}, \cite{myAPS_2021}. As an example, a coaxially-fed, azimuthal mode converter is proposed. The realistic coaxial feed is widely applicable and can be combined with other cylindrical metasurface devices designed using multimodal wave matrix theory \cite{myTAP_2023}.

\begin{figure}[!t]
\centering
\includegraphics[width=8.8cm]{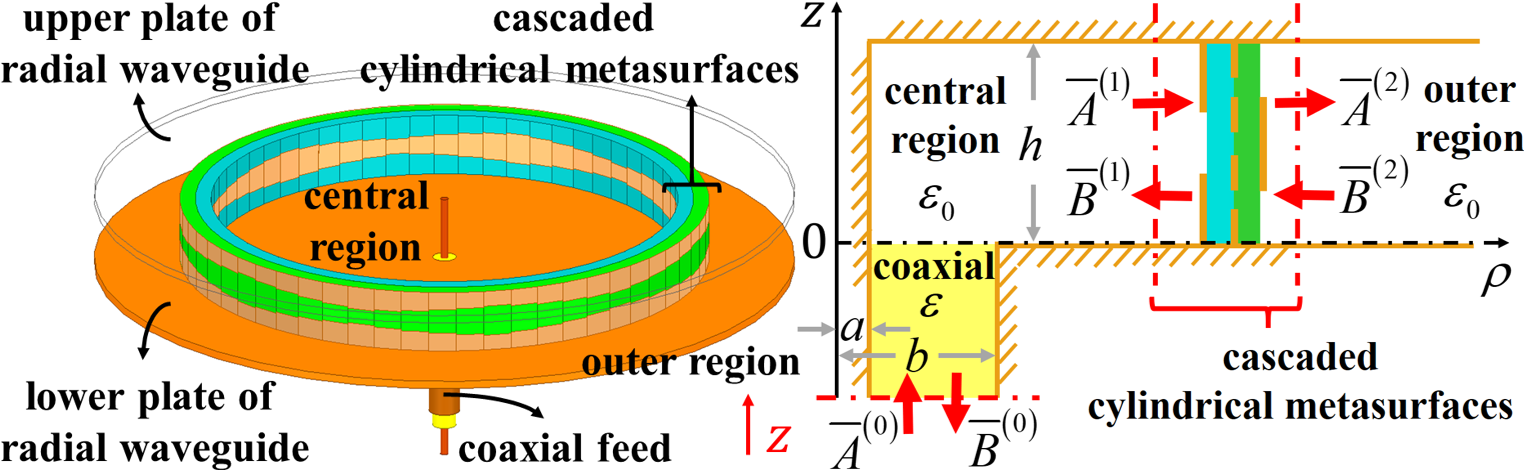}
\caption{Perspective and cross-sectional (a constant $\phi$ plane in cylindrical coordinates) views of cascaded cylindrical metasurfaces excited by a coaxial feed. The inner conductor of the feed touches the upper plate of the waveguide.}
\label{fig_structure}
\end{figure}

\section{Characterization of the Coaxial Feed}
The coaxial feed in Fig. \ref{fig_structure} is chosen such that the inner and outer radii $a$, $b$ and the permittivity $\varepsilon$ allow only TEM$_z$ propagation. In addition, the waveguide height $h$ is selected to ensure all propagating TM$_z$ modes are invariant in $z$ \cite{myAPS_2022}.

\subsection{Overview of the Mode-Matching Technique}
\begin{figure}[!t]
\centering
\includegraphics[width=8.4cm]{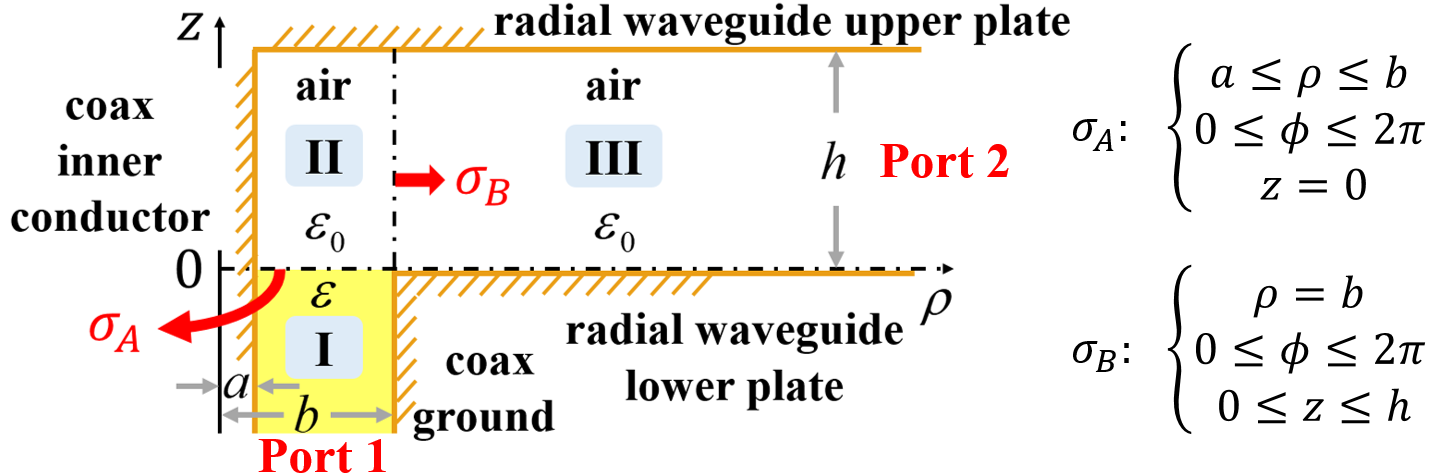}
\caption{A coaxial cable to radial waveguide junction. The junction is rotationally symmetric with respect to the $z$-axis, and is divided into three regions. The interface between region I (dielectric filled) and region II is $\sigma_A$, while the interface between regions II and III is $\sigma_B$.}
\label{fig_regions}
\end{figure}

Our aim is to derive the $S$-matrix of the junction (denoted by subscript $j$) between coaxial feed and radial waveguide:
\begin{equation}
    \begin{bmatrix}
    \bar{B}^{(0)} \\ \bar{A}^{(1)}
    \end{bmatrix}
    =
    \begin{bmatrix}
    \bar{\bar{S}}_{j,11} & \bar{\bar{S}}_{j,12} \\
    \bar{\bar{S}}_{j,21} & \bar{\bar{S}}_{j,22}
    \end{bmatrix}
    \cdot
    \begin{bmatrix}
    \bar{A}^{(0)} \\ \bar{B}^{(1)}
    \end{bmatrix}
,\label{def_S_jx}\end{equation}
\noindent{where port 1 and port 2 of this junction are defined in Fig. \ref{fig_regions}. The $1\times 1$ vectors $\bar{A}^{(0)}$ and $\bar{B}^{(0)}$ (refer to Fig. \ref{fig_structure}) represent the normalized TEM$_z$ waves incident on, and reflected from, the junction, respectively. 
Let us assume there are $N=2M+1$ azimuthal modes (azimuthal orders $m$ from $+M$ to $-M$) in the radial waveguide. 
The $N\times 1$ vectors $\bar{A}^{(1)}$ and $\bar{B}^{(1)}$ contain outward and inward propagating normalized TM$_z$ azimuthal modes in the central region (see Fig. \ref{fig_structure}), and their entries are arranged in descending azimuthal order \cite{myTAP_2023}, \cite{myAPS_2021}.}

Therefore, the matrices in (\ref{def_S_jx}) bear the following properties:
\begin{itemize}[leftmargin=*]
    \item $\Bar{\Bar{S}}_{j,11}$ is an $1\times 1$  matrix with only one entry $S_{j,11(0,0)}$. The second number in parentheses denotes the incident azimuthal order, while the first number indicates the scattered azimuthal order. The azimuthal order of TEM$_z$ waves is 0.
    \item $\Bar{\Bar{S}}_{j,21}$ is an $N\times 1$ matrix. Due to rotational symmetry of the junction, azimuthal mode conversion is not introduced. Thus, among all the entries $S_{j,21(+M,0)}$, $...$, $S_{j,21(-M,0)}$, only the central one $S_{j,21(0,0)}$ is nonzero.
    \item $\Bar{\Bar{S}}_{j,12}$ is an $1\times N$ matrix. Similarly, among all the entries only the central one $S_{j,12(0,0)}$ is nonzero.
    \item $\Bar{\Bar{S}}_{j,22}$ is $N\times N$, and it is a diagonal matrix again because of the rotational symmetry of the junction.
\end{itemize}

To obtain the scattering properties via the mode-matching technique \cite{Shen_1999}-\cite{Kuhn_1973}, the junction is divided into three regions, as shown in Fig. \ref{fig_regions}. In each region, the total fields are expressed as linear combinations of possible modes with some unknown coefficients that can be solved for by matching tangential fields across the interfaces.
For this structure (Fig. \ref{fig_regions}), it can be shown that no TE$_z$ waves will be generated. In this case, the electromagnetic fields can be derived from the magnetic vector potential $\Bar{A}_\text{potential}=\psi\hat{z}$. The electric and magnetic fields can be readily acquired from \cite{Harrington_book}:
\begin{equation}
    \bar{H} = \frac{1}{\mu_0}(\nabla\psi\times\hat{z}),\quad \bar{E} = \frac{1}{j\omega\mu_0\varepsilon_\text{medium}}\nabla\times(\nabla\psi\times\hat{z})
.\label{eq_EH_field}\end{equation}

\subsection{Azimuthally Symmetric Mode ($m=0$) Case}
For simplicity, let us first consider the case where all fields are azimuthally symmetric. Assume that the incident wave comes from port 1 (from the coaxial feed toward the junction).
We first define an $R_m$ function based on Bessel functions of the first kind $J_m$ and second kind $Y_m$ of order $m$ \cite{Shen_1999}, \cite{Marcuvitz_book}:
\begin{equation}
    R_m(x_1, x_2) \triangleq Y_m(x_2)J_m(x_1) - J_m(x_2)Y_m(x_1)
.\label{eq_R_fx}
\end{equation}
\noindent{The potential functions of the scattered modes in region I (the coaxial cable region) are,}
\begin{equation}
\psi^\text{I}_{n_1} \triangleq
    \begin{cases}
    e^{+jk_{zn_1}z}R_0(k_{\rho n_1}\rho, k_{\rho n_1}a) 
    &, n_1 = 1, 2, ... \\
    e^{+jkz}\ln\rho &, n_1 = 0
    \end{cases}
\label{eq_modesI_0}
\end{equation}
\noindent{where $k=\omega\sqrt{\mu_0\varepsilon}$, $k_{\rho n_1}$ is the $n_1$-th value that satisfies the dispersion equation $R_0(k_{\rho n_1}b, k_{\rho n_1}a)=0$, and $k_{zn_1}^2=k^2-k_{\rho n_1}^2$. In (\ref{eq_modesI_0}), $n_1=0$ corresponds to the TEM$_z$ mode, and the nonzero $n_1$ values represent TM$_z$ modes that decay from the junction discontinuity (denoted by interface $\sigma_A$ shown in Fig. \ref{fig_regions}). The total potential in region I contains an incident term and a linear combination of the scattered terms,}
\begin{equation}
    \psi^\text{I}_\text{total} = A^\text{I}_0 e^{-jkz}\ln\rho + \sum_{n_1 = 0}^{\infty} B^\text{I}_{n_1}\psi^\text{I}_{n_1}
,\label{eq_modetotalI_0}\end{equation}
\noindent{in which $A^\text{I}_0$ is known from the incident TEM$_z$ wave and $B^\text{I}_{n_1}$ are unknown coefficients that need to be solved for.}

Region III is the radial waveguide region, so the potential functions of the scattered modes are given as \cite{Harrington_book}:
\begin{equation}
    \psi^\text{III}_{n_3} = \cos\Big(\frac{n_3\pi}{h}z\Big) H_0^{(2)}(k_{\rho n_3}\rho) \quad, n_3 = 0, 1, 2, ...
\label{eq_modesIII_0}
\end{equation}
\noindent{where $k_{\rho n_3}^2 = k_0^2-(n_3\pi/h)^2 = \omega^2\mu_0\varepsilon_0-(n_3\pi/h)^2$. Apart from the $n_3=0$ case, all the other modes are evanescent from the interface $\sigma_B$ (defined in Fig. \ref{fig_regions}) due to imposed waveguide height. The total potential in region III is given by:}
\begin{equation}
    \psi^\text{III}_\text{total} = \sum_{n_3 = 0}^{\infty} A^\text{III}_{n_3}\psi^\text{III}_{n_3}
.\label{eq_modetotalIII_0}
\end{equation}
\noindent{in which $A^\text{III}_{n_3}$ are unknown coefficients to be determined.}

\begin{figure}[!t]
\centering
\includegraphics[width=8.62cm]{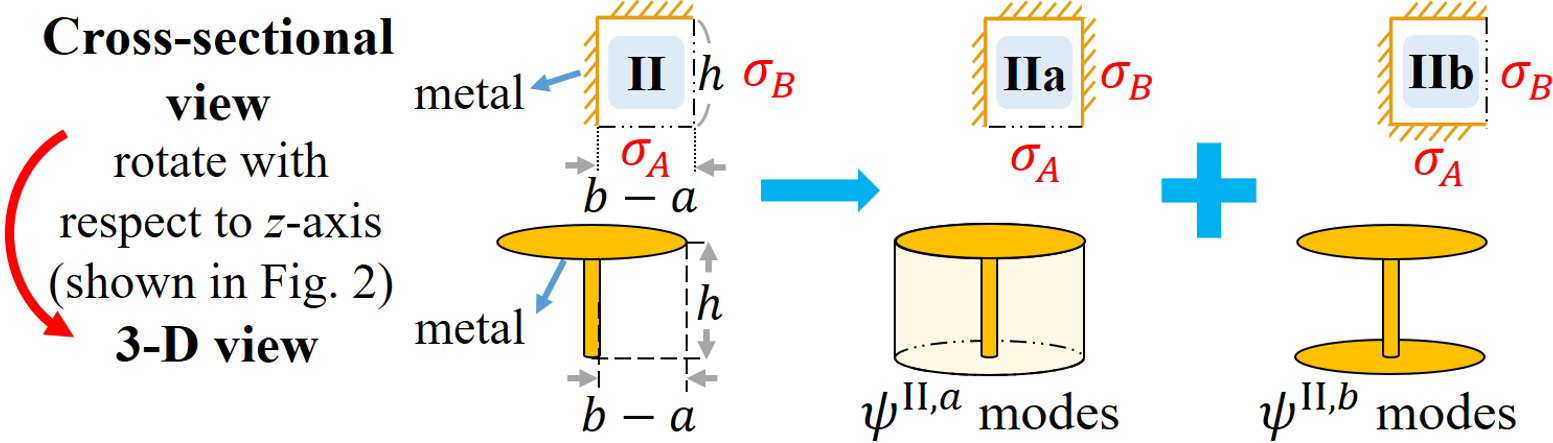}
\caption{An illustration of the resonator method \cite{Shen_1999}, \cite{Kuhn_1973} used to expand the fields in region II.}
\label{fig_regionII}
\end{figure}
In region II, the resonator method, described in \cite{Shen_1999}, \cite{Kuhn_1973} and illustrated in Fig. \ref{fig_regionII}, is applied to express the fields within this region. Based on the Uniqueness Theorem \cite{Harrington_book}, if the tangential electric fields on the interfaces $\sigma_A$ and $\sigma_B$ can be uniquely expanded, the fields within region II can be determined. First, consider the case where only surface $\sigma_B$ is covered with metal. The resulting structure is a coaxial cable short circuited at $z=h$. Analogous to region I, the following $\psi^{\text{II},a}$ potentials are sufficient to represent the tangential electric fields on $\sigma_A$:
\begin{equation}
\psi^\text{II,a}_{n_2} \triangleq
    \begin{cases}
    \cos[k_{zn_2}(z-h)]R_0(k_{\rho n_2}\rho, k_{\rho n_2}a) 
    &, n_2 = 1, 2, ... \\
    \cos[k_0(z-h)]\ln\rho &, n_2 = 0
    \end{cases}
\label{eq_modesIIa_0}
\end{equation}
\noindent{where $k_{\rho n_2}$ is the $n_2$-th value that satisfies the dispersion equation $R_0(k_{\rho n_2}b, k_{\rho n_2}a)=0$, and $k_{zn_2}^2=k_0^2-k_{\rho n_2}^2$.}

On the other hand, consider the alternate case, where surface $\sigma_A$ is replaced with metal, but $\sigma_B$ is not. The resulting structure now resembles the radial waveguide of region III, but with the extra boundary condition of a short circuit at $\rho=a$. Accordingly, the $\psi^{\text{II},b}$ potentials given below fully describe the tangential electric fields on $\sigma_B$: 
\begin{equation}
    \psi^\text{II,b}_{n_2'} = \cos\Big(\frac{n_2'\pi}{h}z\Big) R_0(k_{\rho n_2'}\rho, k_{\rho n_2'}a) \quad, n_2' = 0, 1, 2, ...
\label{eq_modesIIb_0}
\end{equation}
\noindent{where $k_{\rho n_2'}^2 = \omega^2\mu_0\varepsilon_0-(n_2'\pi/h)^2$.
Finally, by superposing these two sets of potentials (\ref{eq_modesIIa_0}) and (\ref{eq_modesIIb_0}), the total potential in region II can be derived,}
\begin{equation}
    \psi^\text{II}_\text{total} = \sum_{n_2 = 0}^{\infty} A^\text{II}_{n_2}\psi^\text{II,a}_{n_2} + \sum_{n_2' = 0}^{\infty} B^\text{II}_{n_2'}\psi^\text{II,b}_{n_2'}
,\label{eq_modetotalII_0}\end{equation}
\noindent{where $A^\text{II}_{n_2}$ and $B^\text{II}_{n_2'}$ are unknown coefficients.}

Now that the total potential in each region has been expressed as a linear combination, the mode-matching technique \cite{Shen_1999}-\cite{Kuhn_1973} could be applied to determine the unknown expansion coefficients.
First, the field quantities are derived by applying (\ref{eq_EH_field}) to the potentials.
At the interface $\sigma_A$, the total tangential electric field in region I, $E^\text{I}_{\rho,\text{total}}$, and that in region II, $E^\text{II}_{\rho,\text{total}}$, are set equal to each other.
Specifically, we use the magnetic field of each $\psi^\text{I}_{n_1}$ mode, $H^\text{I}_{\phi,n_1}$, as a test function, and take the inner product with $E^\text{I}_{\rho,\text{total}}$ and $E^\text{II}_{\rho,\text{total}}$ respectively. Finally, the two expressions are equated,
\begin{equation}
    \langle E^\text{I}_{\rho,\text{total}} \vert H^\text{I}_{\phi,n_1} \rangle
    = \langle E^\text{II}_{\rho,\text{total}} \vert H^\text{I}_{\phi,n_1} \rangle
    \quad \text{for all $n_1$.}
\label{eq_MMT_1}
\end{equation}
\noindent{The total tangential magnetic fields can also be equated and the electric field of each $\psi^{\text{II},a}_{n_2}$ mode used as a test function,} 
\begin{equation}
    \langle E^\text{II,a}_{\rho,n_2} \vert H^\text{I}_{\phi,\text{total}} \rangle
    = \langle E^\text{II,a}_{\rho,n_2} \vert H^\text{II}_{\phi,\text{total}} \rangle
    \quad \text{for all $n_2$}
.\label{eq_MMT_2}
\end{equation}

Similarly, at the interface $\sigma_B$, the total tangential magnetic and electric fields are equated by using $\psi^{\text{II},b}_{n_2'}$ and $\psi^\text{III}_{n_3}$ as test functions, respectively:
\begin{equation}
    \langle E^\text{II,b}_{z,n_2'} \vert H^\text{II}_{\phi,\text{total}} \rangle
    = \langle E^\text{II,b}_{z,n_2'} \vert H^\text{III}_{\phi,\text{total}} \rangle
    \quad \text{for all $n_2'$}
,\label{eq_MMT_3}
\end{equation}
\begin{equation}
    \langle E^\text{II}_{z,\text{total}} \vert H^\text{III}_{\phi,n_3} \rangle
    = \langle E^\text{III}_{z,\text{total}} \vert H^\text{III}_{\phi,n_3} \rangle
    \quad \text{for all $n_3$}
.\label{eq_MMT_4}
\end{equation}
\noindent{These four sets of equations (\ref{eq_MMT_1})-(\ref{eq_MMT_4}) can be cast into a matrix equation and solved simultaneously to obtain the coefficients $B^\text{I}_{n_1}$, $A^\text{II}_{n_2}$, $B^\text{II}_{n_2'}$, and $A^\text{III}_{n_3}$ in terms of the excitation $A^\text{I}_0$.}

Finally, the mode-matching results are used to derive the $S$-matrix of the junction (\ref{def_S_jx}). Let us set the reference surfaces of the $S$-matrix (\ref{def_S_jx}) to $\sigma_A$ and $\sigma_B$ for simplicity. It is convention that an $S$-matrix is unitary when the junction is lossless, and symmetric when the junction is reciprocal. Hence, each mode in the vectors $\Bar{A}^{(0)}$, $\Bar{B}^{(0)}$, $\Bar{A}^{(1)}$ and $\Bar{B}^{(1)}$ given by (\ref{def_S_jx}) needs to be normalized with respect to its power \cite{Pozar_book}-\cite{Kurokawa_1965}. From the analysis of coaxial cables \cite{Pozar_book}, the entries of $\Bar{A}^{(0)}$, $\Bar{B}^{(0)}$ are calculated as $-\sqrt{2\pi\eta\ln(b/a)}A^\text{I}_0$ and $+\sqrt{2\pi\eta\ln(b/a)}B^\text{I}_0$, respectively ($\eta=\sqrt{\mu_0/\varepsilon}$). For waves within the radial waveguide, normalization is discussed in \cite{myTAP_2023}. Therefore, $S$-parameter entries $S_{j,11(0,0)}$ and $S_{j,21(0,0)}$ become:
\begin{equation}
    S_{j,11(0,0)} = -B^\text{I}_0/A^\text{I}_0
,\end{equation}
\begin{equation}
\begin{split}
    &S_{j,21(0,0)} = \\
    &\sqrt{
    2\pi b h\text{Re}
    \Bigg\{
    \frac{j}{\eta_0}
    \frac{H^{(2)'}_0(k_0b)}{H^{(2)}_0(k_0b)}
    \Bigg\}
    }
    \cdot
    \frac{
    (-j\omega\mu_0)A^\text{III}_0 H^{(2)}_0(k_0b)}
    {
    \big(-\sqrt{2\pi\eta\ln(b/a)} A^\text{I}_0 \big)
    }
,\end{split}
\end{equation}
\noindent{where $\eta_0 = \sqrt{\mu_0/\varepsilon_0}$. The matrix $\Bar{\Bar{S}}_{j,11}$ only has one entry $S_{j,11(0,0)}$. Moreover, since in $\Bar{\Bar{S}}_{j,21}$, terms other than $S_{j,21(0,0)}$ are all zero, we have also finished the derivation of $\Bar{\Bar{S}}_{j,21}$. Due to reciprocity, the matrix $\Bar{\Bar{S}}_{j,12}$ can be obtained by simply transposing $\Bar{\Bar{S}}_{j,21}$.}

Additionally, the $S_{j,22(0,0)}$ entry in the $\Bar{\Bar{S}}_{j,22}$ matrix is derived as well. By definition, it is the ratio between the normalized reflected mode and normalized incident mode within the radial waveguide (\ref{def_S_jx}). To consider the case where an incident wave comes from port 2 (see Fig. \ref{fig_regions}), we must remove the excitation term $A^\text{I}_0 e^{-jkz}\ln\rho$ from (\ref{eq_modetotalI_0}), and add a new excitation term $B^\text{III}_0 H^{(1)}_0(k_0\rho)$ to (\ref{eq_modetotalIII_0}). The mode-matching technique (\ref{eq_MMT_1})-(\ref{eq_MMT_4}) is then employed again, and all the unknown coefficients are expressed in terms of $B^\text{III}_0$. By performing this analysis, $S_{j,22(0,0)}$ can be proven to be \cite{myTAP_2023}:
\begin{equation}
    S_{j,22(0,0)} = 
    \big[H_0^{(2)}(k_0b)/H_0^{(1)}(k_0b)\big] \cdot(A^\text{III}_0/B^\text{III}_0)
.\label{eq_Sj22_0}
\end{equation}

\subsection{Higher-Order Azimuthal Mode ($m\neq0$) Case}
Next, let us consider the case where fields have a nonzero azimuthal order $m$. In this case, the potentials and fields possess an $e^{-jm\phi}$ dependence. Although the junction itself is rotationally symmetric, the cylindrical metasurfaces themselves can generate and scatter azimuthally-varying fields.

Since the dimensions of the coaxial cable only allow TEM$_z$ propagation ($m=0$), it is impossible to have a wave with nonzero $m$ incident from port 1, i.e. $\bar{A}^{(0)}$ contains only the TEM$_z$ wave. We only need to consider the case where the wave is incident from port 2 (from the radial waveguide toward the junction). Since $\Bar{\Bar{S}}_{j,11}$, $\Bar{\Bar{S}}_{j,12}$, $\Bar{\Bar{S}}_{j,21}$, and the $S_{j,22(0,0)}$ entry in the diagonal matrix $\bar{\bar{S}}_{j,22}$ are already obtained in the previous section, the goal here is to derive the $S_{j,22(m,m)}$ entry in $\bar{\bar{S}}_{j,22}$. To this end, we first write the potential functions in region I as ($B^\text{I}_{n_1}$ are the unknown expansion coefficients):
\begin{equation}
\psi^\text{I}_{n_1} = e^{+jk_{zn_1}z}R_m(k_{\rho n_1}\rho, k_{\rho n_1}a) e^{-jm\phi}
\quad, n_1 = 1, 2, ... 
\label{eq_modesI_m}
\end{equation}
\begin{equation}
    \psi^\text{I}_\text{total} = \sum_{n_1 = 1}^{\infty} B^\text{I}_{n_1}\psi^\text{I}_{n_1}
,\label{eq_modetotalI_m}\end{equation}
\noindent{where $k_{\rho n_1}$ is the $n_1$-th value that satisfies the dispersion equation $R_m(k_{\rho n_1}b, k_{\rho n_1}a)=0$, and $k_{zn_1}^2=k^2-k_{\rho n_1}^2$. Note that in (\ref{eq_modesI_m}) and (\ref{eq_modetotalI_m}) the TEM$_z$ mode is not excited due to the symmetry of the junction.}

The potential functions in region II can again be found by the resonator method illustrated in Fig. \ref{fig_regionII}. We have one set of coaxial cable modes $\psi^{\text{II},a}$, 
\begin{equation}
\psi^\text{II,a}_{n_2} =
    \cos[k_{zn_2}(z-h)]R_m(k_{\rho n_2}\rho, k_{\rho n_2}a) e^{-jm\phi}
\end{equation}
\noindent{for $n_2 = 1, 2, ...$, where $k_{\rho n_2}$ is the $n_2$-th value that satisfies $R_m(k_{\rho n_2}b, k_{\rho n_2}a)=0$, and $k_{zn_2}^2=k_0^2-k_{\rho n_2}^2$. The other set of modes $\psi^{\text{II},b}$ are similar to those in a radial waveguide,}
\begin{equation}
    \psi^\text{II,b}_{n_2'} = \cos\Big(\frac{n_2'\pi}{h}z\Big) R_m(k_{\rho n_2'}\rho, k_{\rho n_2'}a) e^{-jm\phi} \quad, n_2' = 0, 1, ...
\end{equation}
\noindent{where $k_{\rho n_2'}^2 = \omega^2\mu_0\varepsilon_0-(n_2'\pi/h)^2$. The total potential in region II can be written as the following linear combination due to superposition and uniqueness theorem:}
\begin{equation}
    \psi^\text{II}_\text{total} = \sum_{n_2 = 1}^{\infty} A^\text{II}_{n_2}\psi^\text{II,a}_{n_2} + \sum_{n_2' = 0}^{\infty} B^\text{II}_{n_2'}\psi^\text{II,b}_{n_2'}
,\label{eq_modetotalII_m}\end{equation}
\noindent{with $A^\text{II}_{n_2}$ and $B^\text{II}_{n_2'}$ being the coefficients to be computed.}

Similarly, the potentials for scattered fields in region III are given by the Hankel functions of the second kind of order $m$ which represent outgoing cylindrical waves \cite{Harrington_book},
\begin{equation}
    \psi^\text{III}_{n_3} = \cos\Big(\frac{n_3\pi}{h}z\Big) H_m^{(2)}(k_{\rho n_3}\rho) e^{-jm\phi} \quad, n_3 = 0, 1, ...
\label{eq_modesIII_m}
\end{equation}
\noindent{where $k_{\rho n_3}^2 = \omega^2\mu_0\varepsilon_0-(n_3\pi/h)^2$. The total potential in region III consists of these scattered modes with unknown coefficients $A^\text{III}_{n_3}$, as well as an incident term (incoming cylindrical wave) with a coefficient $B^\text{III}_0$:}
\begin{equation}
    \psi^\text{III}_\text{total} = 
    B^\text{III}_0 H^{(1)}_m(k_0\rho)e^{-jm\phi} + 
    \sum_{n_3 = 0}^{\infty} A^\text{III}_{n_3}\psi^\text{III}_{n_3}
.\end{equation}

Finally, the mode-matching technique, given by (\ref{eq_MMT_1})-(\ref{eq_MMT_4}), is applied to determine the expansion coefficients $B^\text{I}_{n_1}$, $A^\text{II}_{n_2}$, $B^\text{II}_{n_2'}$ and $A^\text{III}_{n_3}$ in terms of the excitation $B^\text{III}_0$. 
Similar to (\ref{eq_Sj22_0}), the $S_{j,22(m,m)}$ entry in the matrix $\bar{\bar{S}}_{j,22}$ is derived as,
\begin{equation}
    S_{j,22(m,m)} = 
    \big[H_m^{(2)}(k_0b)/H_m^{(1)}(k_0b)\big] \cdot(A^\text{III}_0/B^\text{III}_0)
.\label{eq_Sj22_m}
\end{equation}
\noindent{It is worth noting that for $m\neq 0$, the magnitude of $S_{j,22(m,m)}$ must be 1 since these modes are cutoff in the coaxial cable. The $S$-parameter entry (\ref{eq_Sj22_m}) can be computed for all azimuthal modes of interest ($m$ from $-M$ to $+M$). This concludes the derivation of the diagonal matrix $\Bar{\Bar{S}}_{j,22}$. Therefore, the whole $S$-matrix of the junction (\ref{def_S_jx}) has been rigorously determined.} 

\section{Integration of Feed with Cylindrical Metasurfaces}
Now that the coaxial feed has been characterized as an $S$-matrix (\ref{def_S_jx}), the feed can be incorporated into the design of cylindrical metasurfaces. 
The $S$-matrix of cascaded cylindrical metasurfaces can be expressed as:
\begin{equation}
    \begin{bmatrix}
    \bar{B}^{(1)} \\ \bar{A}^{(2)}
    \end{bmatrix}
    =
    \begin{bmatrix}
    \bar{\bar{S}}_{M,11} & \bar{\bar{S}}_{M,12} \\
    \bar{\bar{S}}_{M,21} & \bar{\bar{S}}_{M,22}
    \end{bmatrix}
    \cdot
    \begin{bmatrix}
    \bar{A}^{(1)} \\ \bar{B}^{(2)}
    \end{bmatrix}
.\label{def_S_MTS}\end{equation}
\noindent{As illustrated in Fig. \ref{fig_structure}, (\ref{def_S_MTS}) relates $\bar{A}^{(1)}$ and $\bar{B}^{(1)}$ to the $N\times 1$ vectors $\bar{A}^{(2)}$ and $\bar{B}^{(2)}$, which contain normalized azimuthal modes of the outward and inward propagating $\text{TM}_z$ waves in the outer region. This $S$-matrix (\ref{def_S_MTS}) can be derived through multimodal wave matrix theory \cite{myTAP_2023}, \cite{myAPS_2021}, which is a generalization of traditional network thoery \cite{Collin_book}. It allows for the rigorous analysis and efficient synthesis of cascaded cylindrical metasurfaces, and circumvents several realization difficulties of these structures \cite{myTAP_2023}, \cite{myAPS_2021}.}

In order to capture the interaction between the coaxial feed and the cylindrical metasurfaces, (\ref{def_S_jx}) and (\ref{def_S_MTS}) can be cascaded. 
Consider the coaxial feed shown in Fig. \ref{fig_structure} with excitation $\bar{A}^{(0)}$.
Since the fields generated in the outer region propagate outwardly without reflection, $\Bar{B}^{(2)}$ is a zero vector.
By substituting (\ref{def_S_MTS}) into (\ref{def_S_jx}), the fields within the central region can be obtained in terms of $\bar{A}^{(0)}$:
\begin{equation}
    \Bar{A}^{(1)} = \big{(} \Bar{\Bar{I}} - \Bar{\Bar{S}}_{j,22} \Bar{\Bar{S}}_{M,11}  \big{)}^{-1} \Bar{\Bar{S}}_{j,21} \Bar{A}^{(0)},
    \quad
    \Bar{B}^{(1)} = \Bar{\Bar{S}}_{M,11} \Bar{A}^{(1)}
,\label{field_central}\end{equation}
\noindent{where $\bar{\bar{I}}$ represents an $N\times N$ identity matrix. Using (\ref{field_central}), the total reflection back to the coaxial cable can be written as,}
\begin{equation}
    \Bar{B}^{(0)} = \Big{[}
    \Bar{\Bar{S}}_{j,11} + \Bar{\Bar{S}}_{j,12}\Bar{\Bar{S}}_{M,11} \big{(} \Bar{\Bar{I}} - \Bar{\Bar{S}}_{j,22} \Bar{\Bar{S}}_{M,11}  \big{)}^{-1} \Bar{\Bar{S}}_{j,21} 
    \Big{]}
    \Bar{A}^{(0)}
.\label{field_reflection}\end{equation}
\noindent{The field transmitted to the outer region can be written in terms of normalized azimuthal modes \cite{myTAP_2023} as,}
\begin{equation}
    \Bar{A}^{(2)} = \Bar{\Bar{S}}_{M,21} \big{(} \Bar{\Bar{I}} - \Bar{\Bar{S}}_{j,22} \Bar{\Bar{S}}_{M,11}  \big{)}^{-1} \Bar{\Bar{S}}_{j,21} \Bar{A}^{(0)}
.\label{field_outside}\end{equation}
\noindent{By optimizing the parameters of the cascaded cylindrical metasurfaces, the field in the outer region (\ref{field_outside}) can be tailored to any stipulated field, enabling arbitrary field synthesis.}

As a validation, a coaxially-fed azimuthal mode converter has been designed based on the proposed structure.
It converts an azimuthally symmetric field ($m=0$) mode originating from the coaxial feed to a stipulated field with $e^{-j\phi}$ dependence ($m=1$ mode) in the outer region. 
The operating frequency is set to 10 GHz. The inner and outer conductor radii of the coaxial feed are $a = 0.45$ mm and $b = 1.5$ mm respectively, and the dielectric is chosen to be Teflon with permittivity $\varepsilon = 2.2\varepsilon_0$. 
In addition, the height of the radial waveguide is $h = 5$ mm. 
Following the discussion in \cite{myTAP_2023}, \cite{myTAP_2021}, four metasurface layers, separated by air spacers, are used in design, with radii $\rho_1 = 1.85\lambda$, $\rho_2 = 2.25\lambda$, $\rho_3 = 2.90\lambda$, and $\rho_4 = 3.30\lambda$, where $\lambda$ stands for the operating wavelength. 

\begin{figure}[!t]
\centering
\includegraphics[width=8.8cm]{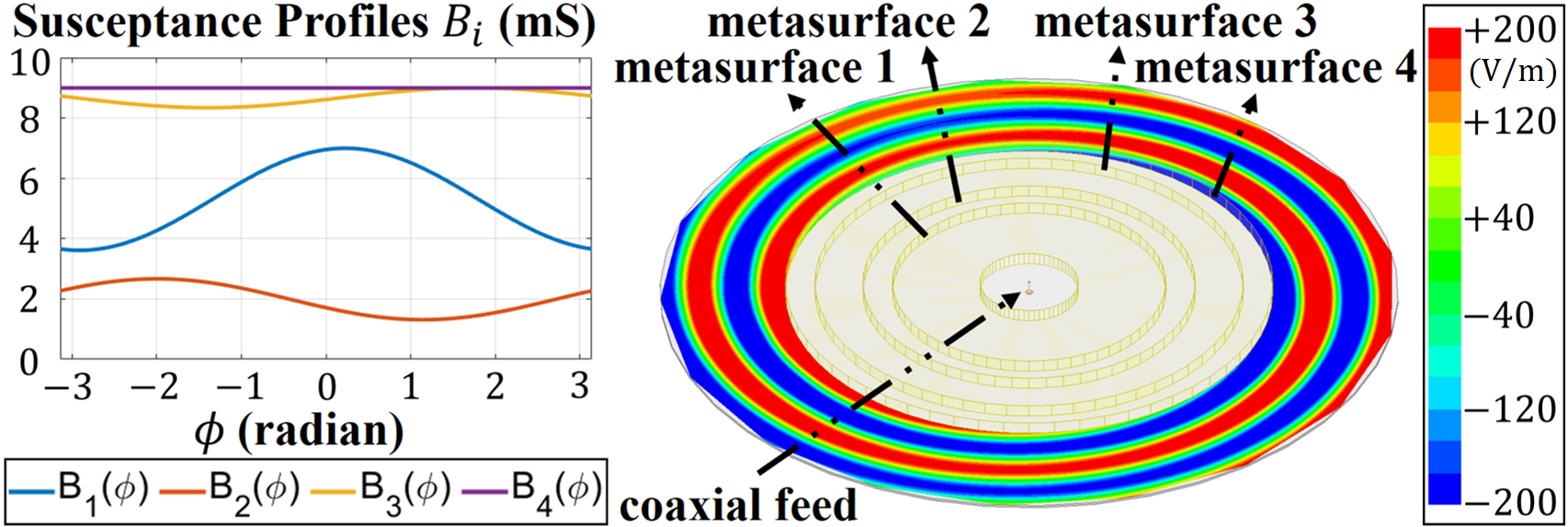}
\caption{The synthesized susceptance for each metasurface layer. Admittance $Y_i(\phi)=jB_i(\phi)$. The real part of the $E_z$ field in the outer region (obtained through full-wave simulation) is also shown.}
\label{fig_Design}
\end{figure}

The $i^\text{th}$ metasurface layer can be modeled by an azimuthally varying admittance profile $Y_i(\phi)$ that relates induced surface electric current density to averaged tangential electric field \cite{Tretyakov_book}.
All $Y_i(\phi)$ profiles are purely imaginary, $Y_i(\phi)=jB_i(\phi)$, for a lossless device.
Finally, the total number of azimuthal modes $N$ considered in the optimization process is set to 31, and the cost function to be minimized is specified as:
\begin{equation}
    \mathcal{C} = \Big{[} 
    \frac{\text{Power of } m = 1 \text{ mode in the outer region}}{\text{Total power in the outer region}} - 1
    \Big{]}^2
.\end{equation}
\noindent{Fig. \ref{fig_Design} shows the optimized, purely imaginary, admittance profiles for all four metasurface layers. Each profile $Y_i(\phi)$ is discretized into 60 unit cells and modeled as penetrable boundary condition in the Ansys HFSS solver.
From the full-wave simulation reported in Fig. \ref{fig_Design}, an $m=1$ mode can be observed in the outer region. 
In the outer region, 93.03\% of the total power is constituted by the $m=1$ mode, while only 1.13\% of the total power is from the $m=0$ mode. The rest of the power is distributed to other azimuthal modes.
The reflection coefficient back at the coaxial cable can be calculated as $\bar{B}^{(0)}[\bar{A}^{(0)}]^{-1}$. This value is $0.781\angle-2.46^\circ$ in our theory and $0.773\angle-2.84^\circ$ in the HFSS simulation. All these results indicate the high accuracy of our proposed characterization of the realistic coaxial feed.}

\section{Conclusion}
Previously, multimodal wave matrix theory has been proposed to design cylindrical metasurfaces that are excited by fictitious current sources. 
In this letter, a realistic coaxial feed is investigated and incorporated into the design process. 
The scattering properties of the feed are accurately computed using the mode-matching technique, and represented as an $S$-matrix. 
The proposed modeling of a realistic coaxial feed can be applied with multimodal wave matrix theory to realize various practical cylindrical-metasurface-based devices. 
Future work includes implementing the admittance profiles with metallic claddings, as discussed extensively in \cite{Tretyakov_book}-\cite{Budhu_MTM_2022}, and minimizing the reflection coefficient of the coaxial feed \cite{Shen_1999}, \cite{Ettorre_2012}.


\newpage

\balance

\end{document}